\documentclass[a4paper,11pt]{article}
\usepackage{pos}

\usepackage{bbm}
\usepackage{slashed}

\title{Renormalization constants of quark bilinear operators in QCD with dynamical up, down, strange and charm quarks}
\ShortTitle{RCs of quark bilinear operators in QCD with dynamical $u$, $d$, $s$ and $c$ quarks}

\manuallySeparateAuthors

\author*[a]{Matteo~Di~Carlo}
\author[b]{, Martha~Constantinou}
\author[c]{, Petros~Dimopoulos}
\author[d]{ and Roberto~Frezzotti}

\author{\ for the Extended Twisted Mass Collaboration (ETMC)}

\affiliation[a]{School of Physics and Astronomy, 
University of Edinburgh,\\ Edinburgh EH9 3FD, United Kingdom}
\affiliation[b]{Department of Physics, Temple University,\\ Philadelphia, PA 19122 - 1801, USA}
\affiliation[c]{Dipartimento  di  Scienze  Matematiche,  Fisiche  e Informatiche, Universit\`a  di  Parma  and  INFN,\\ 
Gruppo Collegato di Parma, Parco Area delle Scienze 7/a (Campus), 43124 Parma, Italy}

\affiliation[d]{Dipartimento di Fisica and INFN, Universit\`a di Roma ``Tor Vergata'',\\
Via della Ricerca Scientifica 1, I-00133 Roma, Italy}

\emailAdd{matteo.dicarlo@ed.ac.uk}

\abstract{We present preliminary results of a calculation of the QCD renormalization constants (RCs) for quark bilinear operators, evaluated non-perturbatively on the lattice in the RI\textquotesingle-MOM scheme. The calculation is performed by using dedicated ensembles with $N_f=4$ degenerate dynamical twisted mass (clover) fermions and the Iwasaki gauge action. A detailed analysis is reported, with emphasis on the control or subtraction of the hadronic contaminations occurring in the lattice estimators of RCs and a check of proper scaling with $a^2$ of the final  results. Such a careful study of systematic errors is the counterpart of the high statistical precision reached by current calculations of RCs in the RI\textquotesingle-MOM scheme and is important in order to quote accurate results in phenomenological applications, such as the computation of quark masses.}

\FullConference{%
 The 38th International Symposium on Lattice Field Theory, LATTICE2021
  26th-30th July, 2021
  Zoom/Gather@Massachusetts Institute of Technology
}

\begin{document}
\maketitle

\section{Introduction}
In this contribution we report on the non-perturbative computation of the renormalization constants (RCs) of quark non-singlet bilinear operators in lattice QCD with Wilson-clover fermions using dedicated gauge ensembles with $N_f=4$ degenerate dynamical quarks produced by the Extended Twisted Mass Collaboration (ETMC)~\cite{ExtendedTwistedMass:2021gbo}. We compute RCs in the RI\textquotesingle-MOM scheme~\cite{Martinelli:1994ty} in which the bilinear operators $O_\Gamma=\overline{q}_1 \Gamma q_2$, with $\Gamma = \{\mathbbm{1},\gamma_5,\gamma_\mu,\gamma_\mu\gamma_5,\sigma_{\mu\nu}\}$ (referred to in the following as $\{S,P,V,A,T\}$, respectively), are renormalized by imposing the condition
\begin{equation}
    Z_q^{-1} Z_{\Gamma} \, \mathrm{Tr}\left.\left[\mathcal{V}_{\Gamma}(p) P_{\Gamma}\right]\right|_{p^2=\mu^2}=1~,
    \label{eq:ren_cond}
\end{equation}
where $Z_{\Gamma}$ is the RC of the operator $O_\Gamma$, $\mathcal{V}_{\Gamma}(p)$ the amputated Green function of the operator between quark and antiquark states with  momentum $p$, which is  projected on its tree-level value with a suitable combination of Dirac matrices $P_{\Gamma}$, and $Z_q$ is the quark field RC.  In this work we adopt the definition of $Z_q$ first proposed in Ref.~\cite{ETM:2010iwh}
\begin{equation}
    Z_q = - \frac{i}{12 N_p}\,\sideset{}{^\prime}{\sum}_{\mu} \mathrm{Tr}\left.\left[\frac{\gamma_\mu}{\tilde{p}_\mu}S_q^{-1}(p)\right]\right|_{p^2=\mu^2}~,
\end{equation}
where $S_q(p)=\sum_x S_q(x,0)e^{-ipx}$ is the quark propagator with momentum $p$ and the sum $\sum_{\mu}'$ is over the $N_p$ non-zero components of the lattice momentum ${a\tilde{p}_\mu \equiv \sin(ap_\mu)}$.
 Throughout the present work we have used twisted mass Wilson-clover fermions in the twisted basis, after which, as usual, the RCs are named.
In the pure gauge sector we use the Iwasaki action~\cite{Iwasaki:1985we} and the clover term coefficient, $c_\mathrm{SW}$, is set to its 1-loop~\cite{Aoki:1998qd} tadpole-boosted value (see Ref.~\cite{Alexandrou:2018egz} for details).
The building block of the calculation of RCs is the lattice estimator of the vertex function $\mathcal{V}_{\Gamma}(p)$, which is constructed in terms of quark propagators with momentum $p$, and is computed for several lattice momenta, different values of the (valence and sea) quark masses and at three different lattice spacings denoted in the following as A, B and C
(see Ref.~\cite{ExtendedTwistedMass:2021gbo} for details on the lattice setup).
Then, we reduce lattice artifacts on $\mathcal{V}_{\Gamma}(p)$ by both applying a "democratic filter" on momenta to reduce hypercubic effects, and by subtracting the perturbative one-loop cut-off effects $\mathcal{O}(g^2 a^\infty)$~\cite{Alexandrou:2015sea}. 
Since RI\textquotesingle-MOM  is a mass-independent scheme, an extrapolation of $\mathcal{V}_{\Gamma}(p)$ to the chiral limit ($m_q\to 0$) is needed to get the RCs. The determination of $Z_P$, in particular, requires more attention due to the chirally divergent behaviour of the vertex function~$\mathcal{V}_P(p)$.
In order to distinguish between $p$-dependent lattice artifacts and the natural dependence on $p^2=\mu^2$ of scale-dependent operators, all RCs are evolved to a common reference scale $\mu_\mathrm{ref}$~\cite{Chetyrkin:1999pq,Gracey:2003yr}. The RC estimates are finally obtained at each lattice spacing by applying an appropriate fit procedure. In this work we will focus mainly on the fit ansatz and results, while more details will be presented in a forthcoming publication~\cite{ETM:RCs_prep}.

\section{Hadronic contaminations in RI-MOM RC estimators}
\label{sec:hadr_cont}
A crucial aspect of this analysis is the careful study of hadronic contaminations in RI-MOM estimators of bilinear RCs. If not treated correctly, they can in fact spoil the determination of RCs. It is well known~\cite{Martinelli:1994ty} that RC estimators in RI-MOM are affected by the contribution of hadronic states which are suppressed as $\mathcal{O}(\Lambda_\mathrm{QCD}^2/p^2)$. These have to be properly identified and removed in order to only get the contribution from external quark states. The evaluation of the RC $Z_P$ of the pseudoscalar density operator requires additional care, since hadronic effects in this case are also chirally divergent as $1/m_\pi^2$. The calculation of $Z_P$ will be briefly described in Sec.~\ref{sec:ZqZpZv} as it is extensively discussed in Ref.~\cite{ExtendedTwistedMass:2021gbo}. The origin of  hadronic contaminations can be understood from the study of the large $p^2$ behaviour ($p^2\gg \Lambda_\mathrm{QCD}^2$) of the 2-fermion Green function
\begin{eqnarray}
    G_{\Gamma}(p) = \int \mathrm{d}^4 x \, \mathrm{d}^4 y \, e^{-ip(x-y)} \langle q_1(x) O_\Gamma(0)\overline{q}_2(y)\rangle 
    \sim S_q(p)\mathcal{V}_{\Gamma}(p) S_q(p) + \sum_{H}  \frac{w_H^{(\Gamma)} \Lambda_\mathrm{QCD}^4}{(p^2 m_H)^2}~.
    \label{eq:GreenFunction}
\end{eqnarray}
The first term corresponds to the contribution from the quark states we want to isolate, while the second is the contribution from intermediate hadronic states. Here $w_H^{(\Gamma)}$ are dimensionless parameters that depend on the specific operator $O_\Gamma$ and, possibly, on the mass of the intermediate hadron state $m_H^2$. 
The hadronic contributions in Eq.~\eqref{eq:GreenFunction} arise from the time orderings where both the quark and antiquark fields are located at time distances  either before or after the operator insertion. It follows that, after amputating $G_{\Gamma}(p)$ with inverse quark propagators ($S_q^{-1}(p)\sim \slashed{p}$) and projecting it on its tree-level value, the hadronic effects in $\mathcal{V}_{\Gamma}(p)$ (and hence in $Z_{\Gamma}$) will be suppressed as $\mathcal{O}(\Lambda_\mathrm{QCD}^2)/p^2$ for all $\Gamma\neq \gamma_5$, and $\mathcal{O}(\Lambda_\mathrm{QCD}^4)/(p^2 m_\pi^2)$ for $\Gamma=\gamma_5$. As a consequence, in our analysis we will  check for the presence of $1/p^2$ contributions in the estimators of the RCs. We stress that the definition of $G_{\Gamma}(p)$ in Eq.~\eqref{eq:GreenFunction} and the corresponding large $p^2$ behaviour on the RHS are peculiar of the RI-MOM scheme in which bilinear vertices have the exceptional kinematics $p_1^2=p_2^2=p^2$ and $q^2=(p_1-p_2)^2=0$. Therefore, other renormalization schemes like RI-SMOM~\cite{Sturm:2009kb} will show a different pattern for hadronic effects. In particular, when using non-exceptional momenta $p_1^2=p_2^2=q^2=p^2$ chiral symmetry breaking  and other infrared effects are expected to be  more suppressed than in RI-MOM. 
An extended discussion on  hadronic contaminations in the RI-MOM scheme is given in Ref.~\cite{ExtendedTwistedMass:2021gbo}, and will be extended in Ref.~\cite{ETM:RCs_prep}, while a study of such effects in the RI-SMOM scheme will be part of future investigations.

\section{Determination of $Z_q$, $Z_P$ and $Z_V$ in RI\textquotesingle-MOM}
\label{sec:ZqZpZv}

In this Section we describe the determination of the RCs of the quark field ($Z_q$), of the pseudoscalar density ($Z_P$), and of the (local) vector current ($Z_V$) obtained in the RI\textquotesingle-MOM scheme.

\begin{figure}[!tb]
    \centering
    \begin{minipage}{0.49\textwidth}
    \centering
    \includegraphics[height=.23\textheight]{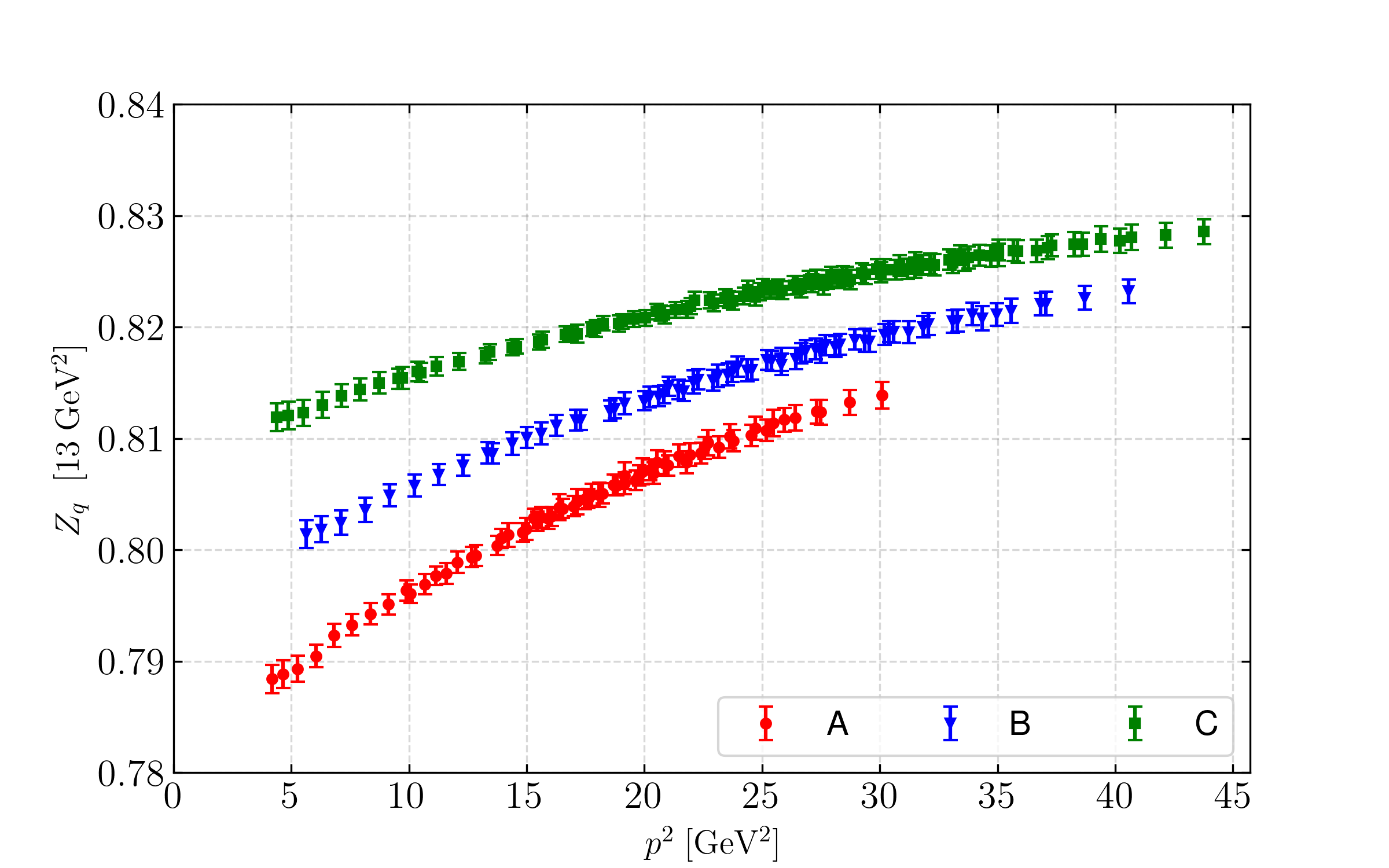}
    \caption{Dependence of $Z_q[13\,\mathrm{GeV}^2]$ on the momentum $p^2$ (in physical units) for the three lattice spacings 
    $a\simeq\{0.0938, 0.0807, 0.0690\}~\mathrm{fm}$, labelled as A, B and C, respectively.}
    \label{fig:Zq}
    \end{minipage}%
    \hfill
    \begin{minipage}{0.48\textwidth}
    \centering
    \includegraphics[height=.19\textheight]{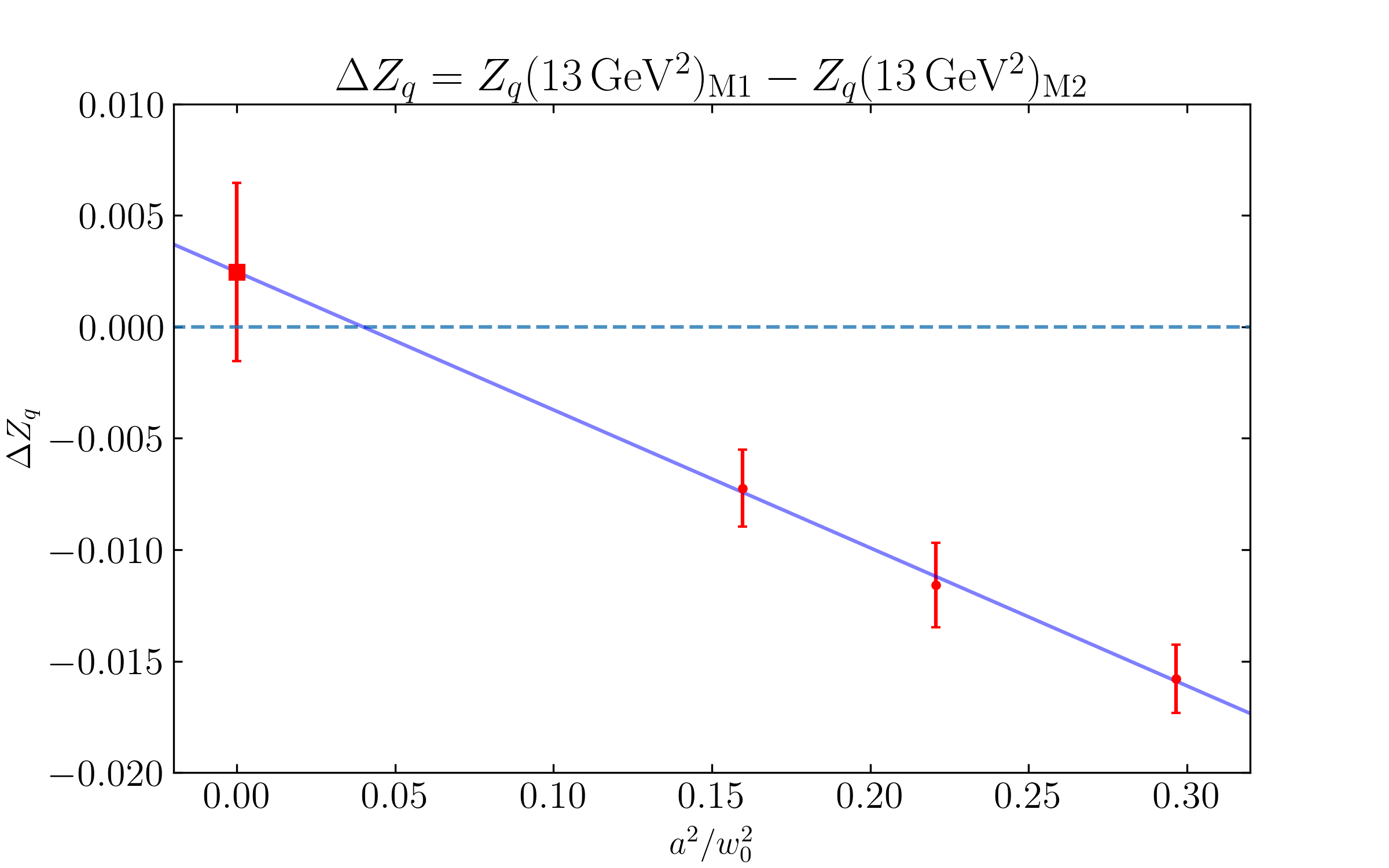}
    \caption{ Scaling of the difference $\Delta Z_q[\textsc{M1}]-Z_q[\textsc{M2}]$ with $(a/w_0)^2$ evaluated at $\mu_\mathrm{ref}^2=13\,\mathrm{GeV^2}$.}
    \label{fig:Zq_scal}
    \end{minipage}
\end{figure}

\paragraph{Quark field renormalization}
The simplest quantity to determine is $Z_q$, as it only depends on the inverse quark propagator and hence does not suffer from hadronic contaminations. Moreover, the dependence of $Z_q$ on the (valence and sea) quark mass is tiny, thus justifying the use of a constant fit ansatz in  the chiral extrapolation. The $p^2$-dependence of the data after the evolution to a common reference scale of $\mu_\mathrm{ref}^2=13\,\mathrm{GeV}^2$ is shown in Fig.~\ref{fig:Zq} for the three lattice spacings.
The final value of $Z_q$ is then obtained using two different methods.  The first method~(\textsc{M1}) consists in fitting the $Z_q(\mu^2_\mathrm{ref})$ data linearly
in $p^2$ in the range $p^2\in[10-16]~\mathrm{GeV}^2$ with the aim of removing $\mathcal{O}(a^2p^2)$  cut-off effects, while in the second method~(\textsc{M2})
the same data are fitted to a constant~\cite{ETM:2010iwh,EuropeanTwistedMass:2014osg}. In principle, \textsc{M2} is by construction much less sensitive than \textsc{M1} to small residual higher order perturbative effects or  hadronic contaminations but leaves some $\mathrm{O}(a^2)$ artifacts in the RC estimates. In the absence of such systematics, the two methods should yield compatible results for renormalized matrix elements in the continuum limit. This can be tested, for example, by studying the scaling of the difference $\Delta Z_{\Gamma} = Z_{\Gamma}[\textsc{M1}]-Z_{\Gamma}[\textsc{M2}]$ with $a^2$. The value of~$\Delta Z_q$, as expected, results to be well compatible with zero in the limit $a^2\to 0$, as shown in Fig.~\ref{fig:Zq_scal}.

\paragraph{The pseudoscalar density} As mentioned in Sec.~\ref{sec:hadr_cont}, the calculation of $Z_P$ is more delicate due to the pion contamination showing up as a pole $1/m_\pi^2$ in the lattice estimator of the vertex function~$\mathcal{V}_P$. 
In our analysis we adopted a partially quenched (PQ) setup\footnote{At all $\beta$ values we compute propagators with 9 different values of the valence masses $\mu_\mathrm{val}$ (from $\sim 8$ to $\sim$16 times larger than the average up-down quark mass) for each of the four sea mass $\mu_\mathrm{sea}$ used.}. Therefore, exploiting that $[m_\pi^2]_\mathrm{val}\sim \mu_\mathrm{val}$ at LO in PQ chiral perturbation theory~\cite{Sharpe:1997by} we can extrapolate the vertex lattice estimator to $\mu_\mathrm{val}=0$ at fixed $p^2$ with the following fit ansatz
\begin{equation}
    v_P(p,\mu_\mathrm{val},\mu_\mathrm{sea}) = \mathcal{V}_P(p,\mu_\mathrm{sea}) + \frac{K'}{p^2} + \frac{K}{p^2}\frac{1}{\mu_\mathrm{val}} + \frac{K''}{p^2}\mu_\mathrm{val} + \dots
    \label{eq:vP_fit_ansatz}
\end{equation}
where $\mathcal{V}_P(p,\mu_\mathrm{sea})$ is our target vertex, the $K$-coefficients contain in general lattice artifacts and the ellipsis stands for terms suppressed by higher powers of $1/p^2$ as $p^2\to\infty$.
In practice, we noticed that the linear dependence on $\mu_\mathrm{val}$ is mostly due to an $\mathcal{O}(a^2)$ effect and therefore $K''$ was not included in the fit, thus absorbing this cut-off effect in the extrapolated value at $\mu_\mathrm{val}=0$. Since at fixed $p^2$ the target vertex $\mathcal{V}_P$ and the finite hadronic contamination $K'/p^2$ cannot be disentangled, after the chiral extrapolation it is necessary to check the data against both $p^2$-dependent lattice artifacts \textbf{and} $\mathcal{O}(a^0)$ residual hadronic effects scaling as $1/p^2$. As described in Ref.~\cite{ExtendedTwistedMass:2021gbo}, we did this in the case of $Z_P(\mu_\mathrm{ref}^2)$ both by fitting directly the $1/p^2$ pole (in a range of $p^2$ around $p^2=\mu_\mathrm{ref}^2$) and by studying the $a^2$ behaviour of the non-perturbative step scaling function $\Sigma_P(\mu_a^2,\mu_b^2)=Z_P(\mu_a^2)/Z_P(\mu_b^2)$. In the first case we obtained a result compatible with zero for the fit parameter of the $1/p^2$ pole, while in the second case we got perfect agreement of numerical data extrapolated linearly in $a^2$ with the continuum perturbative result of $\Sigma_P$ (computed at N$^3$LO). Both findings suggest that residual systematic uncertainties on $Z_P$ are negligible within our small statistical errors.
The final results for $Z_P$ are obtained by applying the methods \textsc{M1} and \textsc{M2} at two different reference scales and using different $p^2$-ranges for the fits. For further details on the calculation of $Z_P$ we refer to  Ref.~\cite{ExtendedTwistedMass:2021gbo}.

\begin{figure}[!b]
    \centering
    \includegraphics[width=.65\textwidth]{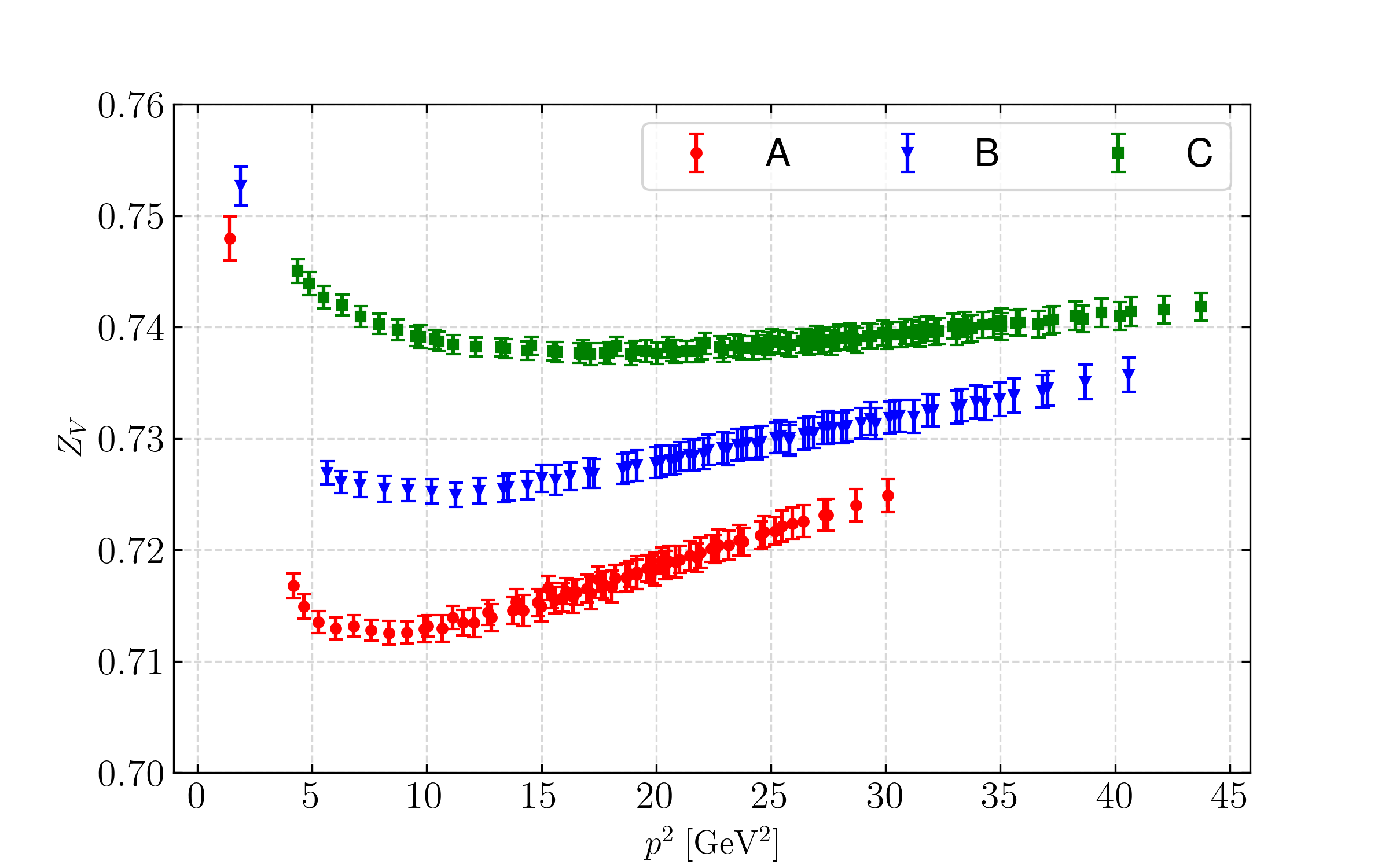}
    \caption{Dependence of $Z_V$ on the momentum $p^2$ (in physical units) for the three lattice spacings 
    ${a\simeq\{0.0938, 0.0807, 0.0690\}~\mathrm{fm}}$, labelled as A, B and C, respectively. }
    \label{fig:ZV}
\end{figure}

\begin{figure}[!hbt]
    \centering
    \begin{minipage}{0.49\textwidth}
    \centering
    \includegraphics[height=.25\textheight]{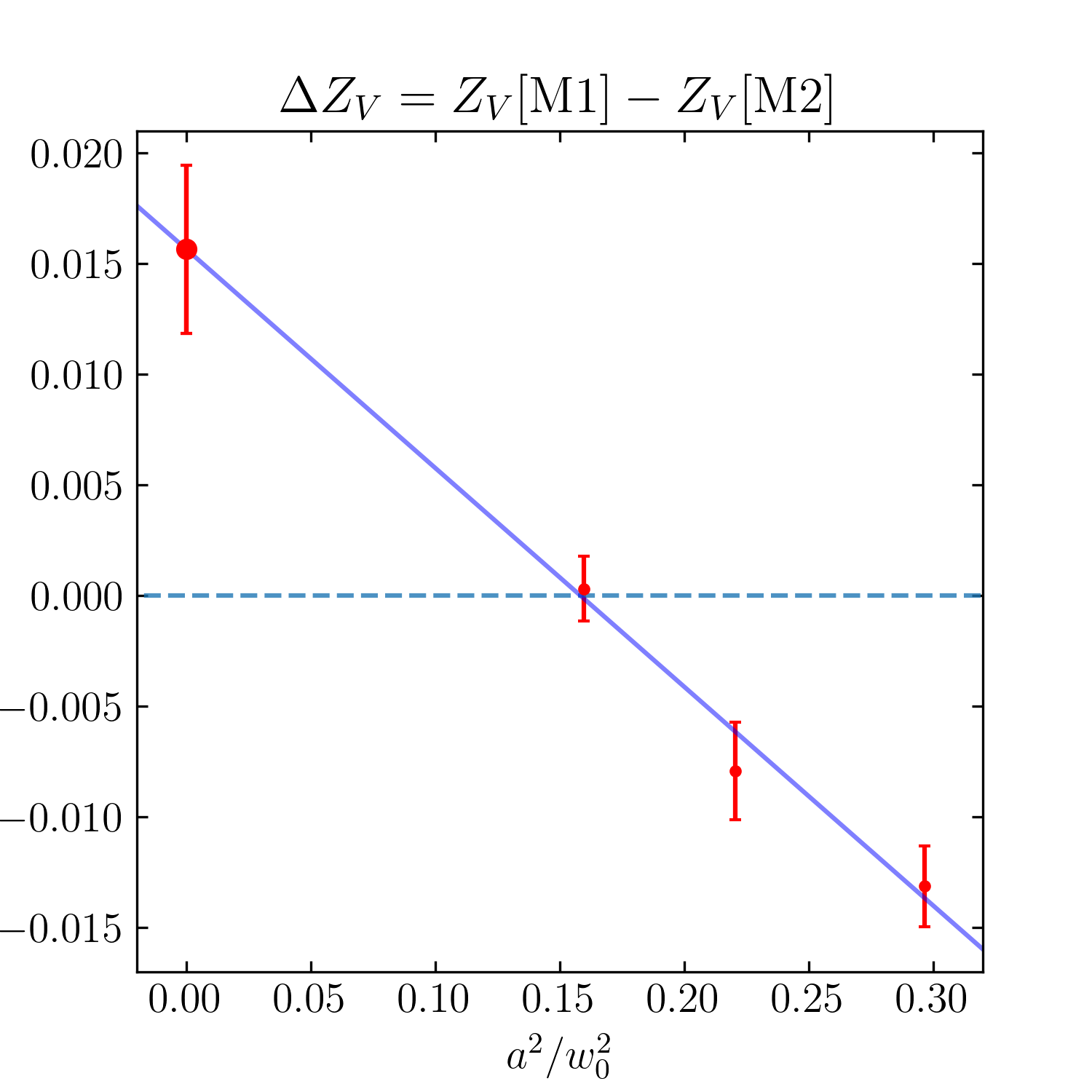}
    \caption{Scaling of the difference $\Delta Z_V = Z_V[\textsc{M1}]-Z_V[\textsc{M2}]$ with $(a/w_0)^2$.}
    \label{fig:ZV_scaling}
    \end{minipage}%
    \hfill
    \begin{minipage}{0.49\textwidth}
    \centering
    \includegraphics[height=.25\textheight]{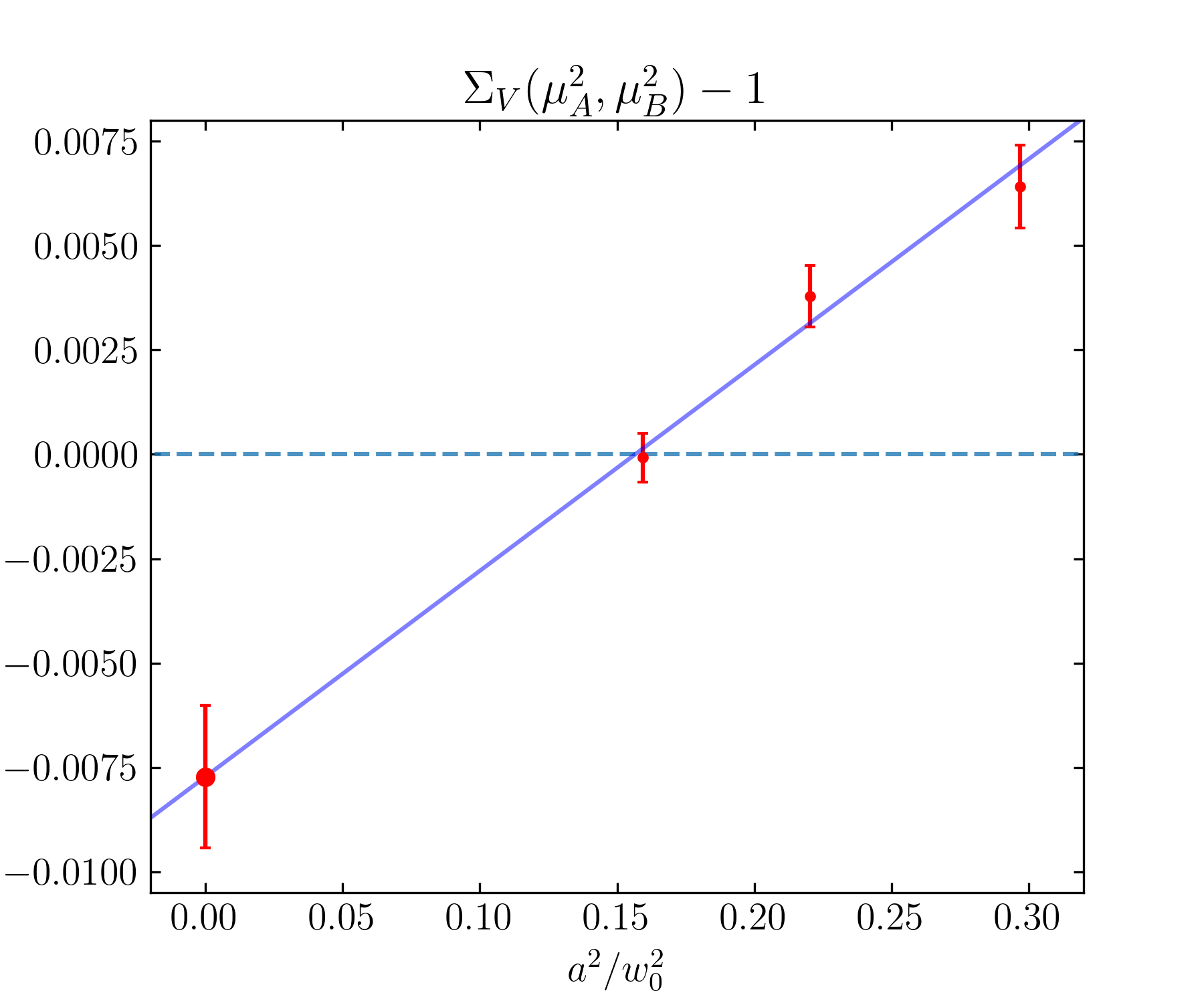}
    \caption{ Scaling of the function $\Sigma_V(21.5\,\mathrm{GeV}^2,14.3\,\mathrm{GeV}^2)-1$  with $(a/w_0)^2$.}
    \label{fig:ZV_stepfunc}
    \end{minipage}
\end{figure}

\paragraph{The local vector current}
We now turn to compute the RC for the local vector current, $Z_V$. In this case, unlike the determination of $Z_P$, the chiral extrapolation is easier due to the absence of chirally divergent effects and thanks to the tiny linear dependence on $\mu_\mathrm{val}$, that allows us to safely neglect the linear term in the chiral fit ansatz. 
However, also in this case one has to take into account the contribution of residual $1/p^2$ hadronic effects, which turn out to be significant for $Z_V$ even at relatively high values of $p^2$. In Fig.~\ref{fig:ZV} we show the $p^2$-dependence of $Z_V$ at three different values of the lattice spacing 
$a\simeq\{0.0938, 0.0807, 0.0690\}~\mathrm{fm}$.
The impact of the $\sim1/p^2$ hadronic effects can be seen very clearly by studying the scaling of $\Delta Z_V = Z_V[\textsc{M1}]-Z_V[\textsc{M2}]$ with $a^2$ (the methods \textsc{M1} and \textsc{M2} being applied in the range $p^2\in[18$-$22]~\mathrm{GeV}^2$ where the curves in Fig.~\ref{fig:ZV} look pretty linear in $p^2$ at all three $\beta$'s)  and by comparing the non-perturbative estimate of the step scaling function $\Sigma_V(\mu_a^2,\mu_b^2)$ with its exact value $\Sigma_V=1$. Both tests fail dramatically:  from Fig.~\ref{fig:ZV_scaling} we see that the quantity $\Delta Z_V$ does not scale linearly to zero with $(a/w_0)^2$, while in Fig.~\ref{fig:ZV_stepfunc} the same behaviour is shown for the difference of the lattice step scaling function $\Sigma_V(21.5~\mathrm{GeV}^2,14.3~\mathrm{GeV}^2)$ from unit value.
This means that the methods \textsc{M1} and \textsc{M2} are not adequate in this case to extract the value of $Z_V$. Therefore, we introduce a new method \textsc{M3}, consisting in a fit of the data on large $p^2$ ranges with the following fit ansatz
\begin{equation}
\zeta_V(p^2) = Z_V \ + \ d_2^{(V)}\!\cdot(a^2 p^2)\ +\ \frac{\epsilon_V}{p^2}\,.
\label{eq:ZV_M3}
\end{equation}
Here $\zeta_V(p^2)$ is the lattice estimator of the RC (corresponding to the curves in Fig.~\ref{fig:ZV}), while $Z_V$ is our target RC, which is free from leading cut-off effects and hadronic contaminations. The method \textsc{M3} is applied in the ranges $p^2\in[4$-$20]~\mathrm{GeV}^2$, $p^2\in[4$-$25]~\mathrm{GeV}^2$ and $p^2\in[4$-$30]~\mathrm{GeV}^2$ for the lattice spacings A, B and C, respectively. The choice of different upper bounds is motivated by the fact that higher order cut-off effects start contributing at different values of physical $p^2$ depending on the lattice spacing. We also checked that the fit is not sensitive to small variations of the lower bound. The results of the fit parameters obtained with the \textsc{M3} method are reported in Table~\ref{tab:ZV_M3}, where the second uncertainty quoted for $Z_V$ is a conservative systematic error.
\begin{table}[!hbt]
    \centering
\begin{minipage}{0.55\textwidth}
\centering
    \begin{tabular}{|c|c|c|c|}
\hline
$\beta$ &$Z_V$ & $d_2^{(V)}$ & $\epsilon_V \ [\mathrm{GeV^2}]$\\
\hline
\hline
1.726 & 0.6989\,(15){(10)} & 0.0038\,(3) & 0.057\,(6)\\
1.778 & 0.7148\,(22){(10)} & 0.0032\,(4) & 0.051\,(11)\\
1.836 & 0.7306\,(13){(10)} & 0.0018\,(4) & 0.061\,(6)\\
\hline
\end{tabular}
    \caption{Results of the fit on $\zeta_V(p^2)$ using the \textsc{M3} fit ansatz of Eq.~\eqref{eq:ZV_M3}.}
    \label{tab:ZV_M3}
\end{minipage}
\hfill
\begin{minipage}{0.40\textwidth}
\centering
    \begin{tabular}{|c|c|}
\hline
$\beta$ &$Z_V\,\mathrm{[WTI]}$ \\
\hline
\hline
1.726 & 0.6960\,(7)\\
1.778 & 0.7131\,(6)\\
1.836 & 0.7310\,(5)\\
\hline
\end{tabular}
    \caption{Results of $Z_V$, in the chiral limit, obtained from the WTI in Eq.~\eqref{eq:ZV_WTI}.}
    \label{tab:ZV_WTI}
\end{minipage}
\end{table}

\begin{figure}
    \centering
    \includegraphics[width=.55\textwidth]{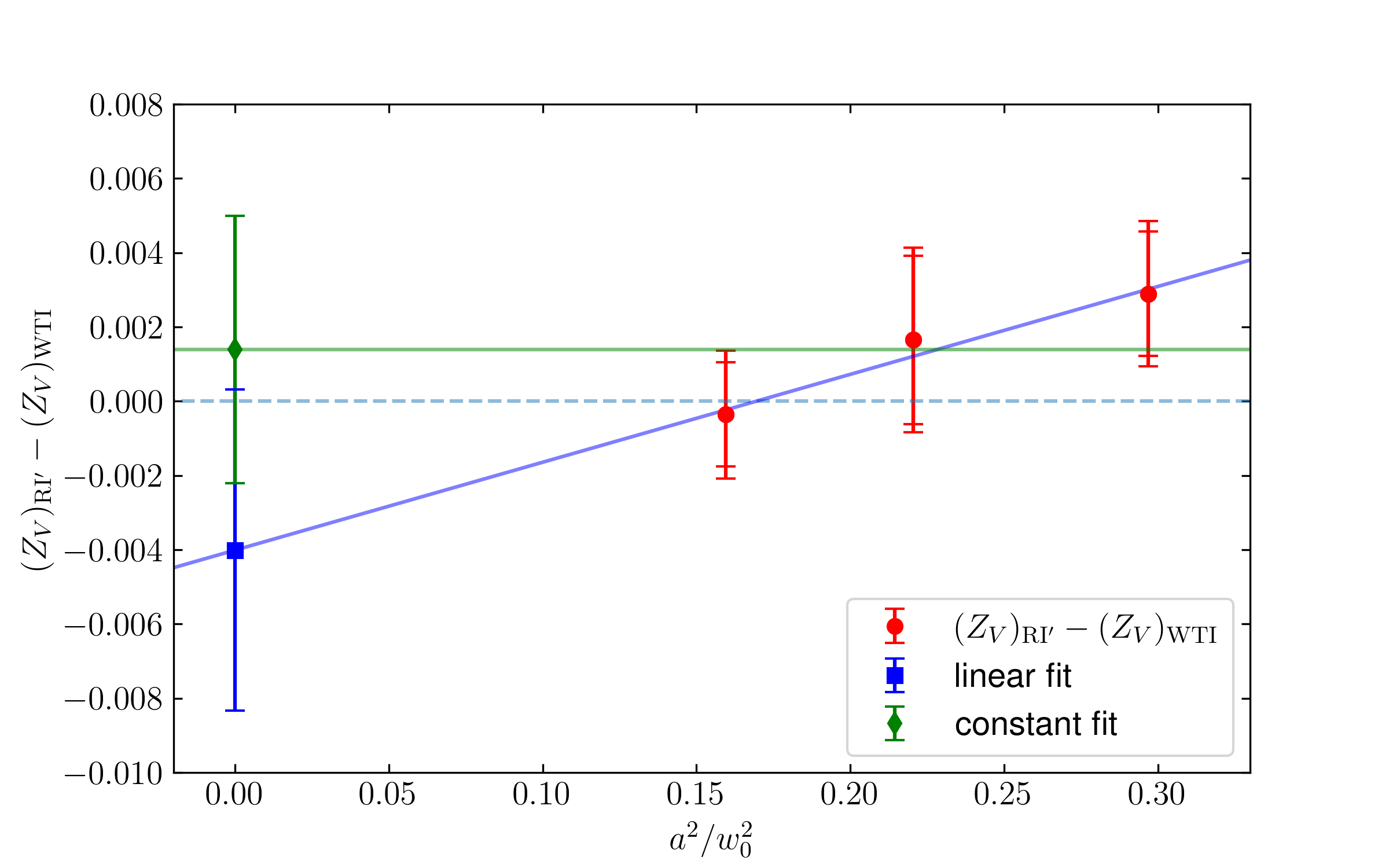}
    \caption{Scaling of the difference $(Z_V)_\text{RI\textquotesingle}-(Z_V)_\text{WTI}$ versus $a^2/w_0^2$. The larger errorbars on the data represent the statistical+systematic uncertainty propagated from the results in Table~\ref{tab:ZV_M3}. A linear fit (blue) and a constant fit (green) to the data are also reported.}
    \label{fig:ZV_RI_WTI}
\end{figure}

\noindent
The RC $Z_V$ can be obtained also in alternative and completely independent ways from the study of hadronic correlators involving the vector current. In particular, we can extract $Z_V$ from the PCAC Ward-Takahashi identity (WTI) for twisted mass fermions, namely
\begin{eqnarray}
    Z_V \sum_\mathbf{x} \langle \tilde{\partial}_0 V_0(t,\mathbf{x}) P^\dagger(0)\rangle^{(\chi)} = (\mu_1+\mu_2) \, \sum_\mathbf{x} \langle P(t,\mathbf{x}) P^\dagger(0)\rangle^{(\chi)}\,,
    \label{eq:ZV_WTI}
\end{eqnarray}
where $\tilde{\partial}_0$ is the symmetric lattice time derivative,  $P$ is the pseudoscalar density and the suffix $(\chi)$ denotes that the matrix elements are computed using twisted mass fermions in the \emph{twisted} basis, in which the renormalized (physical) axial current can be written as $(A_R)_\mu=Z_V A_\mu^{(q)}= - i Z_V V_\mu^{(\chi)}$. The results obtained with this method using the same $N_f=4$ gauge ensembles are reported in Table~\ref{tab:ZV_WTI}. A remarkable advantage of this approach is that the correlators in Eq.~\eqref{eq:ZV_WTI} can be computed very precisely and $Z_V$ can be extracted without any systematic effect other than $\mathcal{O}(a^2)$ artifacts.
The values of $Z_V$ obtained from the WTI can then be used to check the reliability of the RI-MOM results in Tab.~\ref{tab:ZV_M3}.
The two determinations, if hadronic contaminations are under control, should in fact only differ by $\mathcal{O}(a^2)$ discretization effects. 
The scaling of the difference $Z_V[\text{RI\textquotesingle}]-Z_V[\text{WTI}]$ is shown in Fig.~\ref{fig:ZV_RI_WTI}. The larger errorbars on the data in the figure represent the  statistical+systematic uncertainty propagated from the results in Table~\ref{tab:ZV_M3}. We show the results of a constant fit and a linear fit to the data (using the larger errorbars), both yielding an intercept well compatible with zero. Notice that all  three points at finite lattice spacing are compatible to zero within 1-2$\sigma$.
The calculation of $Z_V$ through the WTI is certainly preferable, since it allows one to reach a very high level of precision.
However, we believe that the calculation of $Z_V$ in the RI\textquotesingle-MOM scheme is valuable and instructive, since the good agreement between the two approaches highlights the reliability of our analysis procedure and shows that the use of RI-MOM  does not prevent precise determinations of RCs at the per mille level, once hadronic contaminations are properly treated.

\section{Evaluation of the other renormalization constants}
For the determination of the RCs of the axial current ($Z_A$) and of the scalar density ($Z_S$) we have employed a different approach, since our safety checks against hadronic contaminations failed in both cases and we did not have alternative determinations as precise as $Z_V$[WTI] to compare with. Therefore our approach here consists in an indirect evaluation based on a combination of results obtained  in  RI\textquotesingle-MOM and using alternative hadronic methods.
\paragraph{The axial  current} We determine $Z_A$ as the combination of $(Z_A/Z_V)[\text{RI\textquotesingle}]$ obtained in the RI\textquotesingle-MOM scheme and $Z_V$[WTI] from Ward identities. In fact, the lattice estimator of the ratio $Z_A/Z_V$ shows a suppression of the hadronic contaminations compared to $Z_A$ alone. We then follow the same procedure described in Sec.~\ref{sec:ZqZpZv} for $Z_V$ and compute the ratio $Z_A/Z_V$ in the RI\textquotesingle-MOM scheme applying the method \textsc{M3} to the lattice estimator $\zeta_{A/V}(p^2)$ in the ranges $p^2\in[4\text{-}20]~\mathrm{GeV}^2$, $p^2\in[4\text{-}25]~\mathrm{GeV}^2$ and $p^2\in[4\text{-}30]~\mathrm{GeV}^2$ for the lattice spacings A, B and C, respectively. The results of the fit are reported in the second column of Table~\ref{tab:ZAV_RI_tmOS}. As a safety check, these results can be compared with an alternative and independent determination of $Z_A/Z_V$ that only makes use of hadronic matrix elements and exploits the properties of two variants of the twisted-mass fermion action~\cite{ETM:2010iwh}. In this case, the ratio $Z_A/Z_V$ is extracted by expressing the renormalized physical matrix element of the axial current $\langle 0 | (A_0)_R |\pi\rangle$ in terms of matrix elements of bare operators regularized using either the twisted mass (tm) or Osterwalder-Seiler (OS) valence quarks\footnote{The two regularizations once expressed in the \emph{physical} quark basis differ by the values of the Wilson parameters of the $u,d$ quark fields, namely $r_u=-r_d=1$ for tm fermions and $r_u=r_d=1$ for OS fermions. See Ref.~\cite{ETM:2010iwh} for further details.}:
\begin{equation}
   \langle 0 | (A_0)_R |\pi\rangle =  Z_V \langle 0 | A_0 | \pi\rangle^\mathrm{tm} = Z_A \langle 0 | A_0 | \pi\rangle^\mathrm{OS} + \mathcal{O}(a^2)~.
\end{equation}
The two matrix elements differ only by $\mathcal{O}(a^2)$ effects at maximal twist and therefore $Z_A/Z_V$ is readily obtained from their ratio. The results obtained with this method using the same $N_f=4$ gauge ensembles are reported in the third column of Table~\ref{tab:ZAV_RI_tmOS}. The difference  $(Z_A/Z_V)\text{[RI\textquotesingle]}-(Z_A/Z_V)\text{[tm/OS]}$ has a very good scaling with $(a/w_0)^2$ also in this case. Notice that currently the precision on $Z_A/Z_V$ obtained in this way is much lower than for $Z_V$[WTI]. However, this can be easily improved in the future by increasing the statistics. Finally, we extract  $Z_A$ by combining the ratio $Z_A/Z_V$ from RI\textquotesingle-MOM and $Z_V$ from WTI, namely as $Z_A = (Z_A/Z_V)[\text{RI\textquotesingle}]\times Z_V[\text{WTI}]$\,. This procedure yields an overall uncertainty of $\mathcal{O}(0.1\%)$ on $Z_A$, mainly driven by the RI\textquotesingle-MOM determination $Z_A/Z_V[\text{RI\textquotesingle}]$.

\begin{table}[!htb]
    \centering
    \begin{tabular}{|c|c|c|}
\hline
$\beta$ & $Z_A/Z_V \ [\mathrm{RI\textquotesingle}]$ & $Z_A/Z_V \ [\mathrm{tm/OS}]$\\
\hline
\hline
1.726 & 1.0599\,(13) & 1.036\,(7)\\
1.778 & 1.0563\,(11) & 1.035\,(5)\\
1.836 & 1.0524\,(8)  & 1.040\,(4)\\
\hline
\end{tabular}
    \caption{Results for the ratio $Z_A/Z_V$ obtained in the RI\textquotesingle-MOM scheme using the \textsc{M3} method (second column) and with the alternative hadronic method through chiral extrapolation (third column).}
    \label{tab:ZAV_RI_tmOS}
\end{table}

\paragraph{The scalar density} The scalar RC $Z_S$ is obtained analogously as a combination of results from RI\textquotesingle-MOM and hadronic methods. In this case, since $Z_P$ is extracted very precisely in the RI\textquotesingle-MOM scheme (see Sec.~\ref{sec:ZqZpZv} and Ref.~\cite{ExtendedTwistedMass:2021gbo}), we extract $Z_S$ by combining such result with the ratio $(Z_S/Z_P)[\text{tm/OS}]$ obtained from the calculation of the hadronic matrix element of the pseudoscalar density $\langle 0|P|\pi\rangle $ in the tm and OS regularizations, namely $Z_S = (Z_S/Z_P)[\text{tm/OS}]\times Z_P[\text{RI\textquotesingle}]$\,. The current precision on $Z_S$ is $\mathcal{O}(1\%)$ and it is dominated by the RI\textquotesingle-MOM determination of $Z_P$.

\paragraph{The tensor operator} The calculation of $Z_T$ is still in progress and its accuracy has to be established. The complete analysis will be presented in a forthcoming publication~\cite{ETM:RCs_prep}.

\section{Conclusion}
In this work we have presented an overview and some preliminary results about the calculation of RCs of quark bilinear operators performed using $N_f=4$ ETMC gauge ensembles~\cite{ExtendedTwistedMass:2021gbo}. 
Since many current lattice calculations require high precision level, systematic uncertainties in the determination of RCs become relevant and must be kept under control. For this reason, in our analysis we carry on a careful study of the hadronic contaminations affecting lattice estimators of RCs, evaluated  non-perturbatively in the RI\textquotesingle-MOM scheme. The different impact of such hadronic effects on the  Green functions of quark bilinear operators suggests different numerical strategies to extract the various RCs. In the case of significant hadronic contaminations, these have been identified and subtracted from the data, as it has been shown in the case of $Z_V$. Scale-independent combinations of RCs can also be computed using alternative methods based on ratios of suitable hadronic matrix elements. Looking at the $a^2$-scaling of the difference of the results from the hadronic and  RI\textquotesingle-MOM methods we get a conservative upper bound on the residual systematic errors. 

\begin{acknowledgments}

M.D.C.~is supported by UK STFC grant ST/P000630/1. P.D.~acknowledges support from the European Union Horizon 2020 research and innovation programme under the Marie Sk\l{}odowska-Curie grant agreement No.~813942 (EuroPLEx) and acknowledges also support from INFN under the research project INFN-QCDLAT. R.F.~acknowledges partial support from
the University of Tor Vergata through the Grant “Strong Interactions: from Lattice QCD
to Strings, Branes and Holography” within the Excellence Scheme “Beyond the Borders”.
\end{acknowledgments}

{
\bibliographystyle{JHEP}
\bibliography{bibliography.bib}
}
\end{document}